\documentclass [12pt] {article}
\usepackage{graphicx}
\usepackage{epsfig}
\def\lsim{\raise0.3ex\hbox{$<$\kern-0.75em\raise-1.1ex\hbox{$\sim$}}}
\def\gsim{\raise0.3ex\hbox{$>$\kern-0.75em\raise-1.1ex\hbox{$\sim$}}}

\begin {document}

\begin{center}
{\Large {\bf BARYON NUMBER TRANSFER}} \\
\vskip 0.3cm
{\Large {\bf IN HIGH ENERGY $hp$ COLLISIONS}} \\

\vskip 1.5 truecm
{\bf F. Bopp and Yu. M. Shabelski$^1$}\\
\vskip 0.5 truecm
Siegen Univ., Germany \\
E-mail: Bopp@physik.uni-siegen.de
\end{center}
\vskip 1.5 truecm
\begin{center}
{\bf ABSTRACT}
\end{center}

The process of baryon number transfer due to string junction propagation
in rapidity is considered. It has a significant effect
in the net baryon production in $pp$ collisions at mid-rapidities and
an even larger effect in the forward hemisphere in the cases
of $\pi p$ and $\gamma p$ interactions. The results of numerical
calculations in the framework of the Quark-Gluon String Model are in
reasonable agreement with the data with the same parameter values for
different energies.

\vskip 1.5 truecm

\noindent $^1$Permanent address: Petersburg Nuclear Physics Institute,
Gatchina, St.Petersburg, Russia
E-mail: shabelsk@thd.pnpi.spb.ru

\newpage
\pagestyle{plain}
\noindent{\bf 1. INTRODUCTION}
\vskip 0.5 truecm

The Quark--Gluon String Model (QGSM) and the Dual Parton Model (DPM) are
based on the Dual Topological Unitarization (DTU) and describe quite
reasonably many
features of high energy production processes, including the inclusive
spectra of different secondary hadrons, their multiplicities,
KNO--distributions,
etc., both in hadron--nucleon and hadron--nucleus collisions
\cite{KTM,2r,KTMS,Sh,Sh1}.
High energy interactions are considered as proceeding via the exchange
of one or several pomerons and all elastic and inelastic processes
result from cutting through or between pomerons \cite{AGK}. The
possibility of exchanging a different number of pomerons introduces
absorptive corrections to the cross sections which are in
agreement with the experimental data on production of hadrons consisting
of light quarks. Inclusive spectra of hadrons are related to the
corresponding fragmentation functions of quarks and diquarks, which
are constructed using the reggeon counting rules \cite{Kai}.

In the present paper we discuss the processes connected with the transfer
of baryon charge over long rapidity distances. In the string models
baryons are considered as configurations consisting of three strings
attached to three valence quarks and connected in a point
called "string junction" \cite{IOT,RV}.  Thus the string-junction has a
nonperturbative origin in QCD.

It is very interesting to understand the role of the string-junction in
the dynamics of high-energy hadronic interactions, in particular in the
processes of baryon number transfer \cite{Khar}. The important results
were obtained in \cite{ACKS}. In this paper we prolong to study this
problem. We find a set of the model parameters which can describe all
experimental data concerning baryon number transfer. Feynman scaling
violation for leading baryons is discussed. We also present
the description of new experimental data.

\vskip 0.2 truecm
\noindent{\bf 2. INCLUSIVE SPECTRA OF SECONDARY HADRONS IN QGSM}
\vskip 0.5 truecm

As mentioned above, high energy hadron--nucleon and hadron--nucleus
interactions
are considered in the QGSM and in DPM as proceeding via the exchange of one
or
several pomerons. Each pomeron corresponds to a cylindrical diagram
(see Fig. 1a), and thus, when cutting a pomeron two showers of secondaries
are produced (Fig. 1b). The inclusive spectrum of secondaries is
determined by the convolution of diquark, valence and sea quark
distributions $u(x,n)$ in the incident particles and the fragmentation
functions $G(z)$ of quarks and diquarks into secondary hadrons.

The diquark and quark distribution functions depend on the number $n$ of cut
pomerons in the considered diagram. In the following we use the formalism of
QGSM.

In the case of a nucleon target the inclusive spectrum of a secondary
hadron $h$ has the form \cite{KTM}:

\begin{equation} \frac{x_E}{\sigma_{inel}}
\frac{d\sigma}{dx} =\sum_{n=1}^{\infty}w_{n}\phi_{n}^{h}(x)\ \ ,
\end{equation}
where $x$ is the Feynman variable $x_F$ and $x_E=2E/\sqrt(s)$

The functions $\phi_{n}^{h}(x)$ determine the contribution of
diagrams with $n$ cut pomerons and $w_{n}$ is the probability of this
process. Here we neglect the contributions of diffraction dissociation
processes which are comparatively small in most of the processes
considered below. It can be accounted for separately \cite{KTM,2r}.

For $pp$ collisions
\begin{equation}
\phi_{pp}^{h}(x) = f_{qq}^{h}(x_{+},n)f_{q}^{h}(x_{-},n) +
f_{q}^{h}(x_{+},n)f_{qq}^{h}(x_{-},n) +
2(n-1)f_{s}^{h}(x_{+},n)f_{s}^{h}(x_{-},n)\ \  ,
\end{equation}

\begin{equation}
x_{\pm} = \frac{1}{2}[\sqrt{4m_{T}^{2}/s+x^{2}}\pm{x}]\ \ ,
\end{equation}
where the transverse mass of the produced hadron
$m_T = \sqrt{m^2 + p^2_T}$
and $f_{qq}$, $f_{q}$ and $f_{s}$ correspond to the contributions of
diquarks, valence and sea quarks respectively. They are determined by
the convolution of the diquark and quark distributions with the
fragmentation functions, e.g.,
\begin{equation}
f_{q}^{h}(x_{+},n) = \int_{x_{+}}^{1} u_{q}(x_{1},n)G_{q}^{h}(x_{+}/x_{1})
dx_{1}\ \ .
\end{equation}
In the case of a meson beam the diquark contributions in Eq. (2) should
be changed by the contribution of valence antiquarks:
\begin{equation}
\phi_{\pi p}^{h}(x) = f_{\bar{q}}^{h}(x_{+},n)f_{q}^{h}(x_{-},n) +
f_{q}^{h}(x_{+},n)f_{qq}^{h}(x_{-},n) +
2(n-1)f_{s}^{h}(x_{+},n)f_{s}^{h}(x_{-},n)\ \  .
\end{equation}

The diquark and quark distributions as well as the fragmentation
functions are determined from Regge intercepts. Their expressions are given
in Appendix 1 of \cite {ACKS}. In the present calculations we use the
same function with only one exception.

The direct fragmentation of the initial baryon into the secondary one
(nucleon or lambda/sigma hyperons) with conservation of the string
junction can go via three different processes (Figs.~2a-2c). Obviously,
in the case of $\Xi$ production only two possibilities exist with
string junction plus either one valence quark and two sea quarks or
three sea quarks. In the case of production of a secondary baryon
having no common quarks with the incident nucleons only the string
junction without valence quarks can contribute (Fig.~2c).

All these contributions are determined by Eqs. similar to Eq. (4)
with the corresponding fragmentation functions given by

\newpage

\begin{equation}
G_{uu}^p = G_{ud}^p = a_N z^{\beta} [v_0\varepsilon (1-z)^2 +
v_q z^{2 - \beta} (1-z) + v_{qq}z^{2.5 - \beta}] \; ,
\end{equation}
\begin{equation}
 G_{ud}^{\Lambda} = a_N z^{\beta}
[v_0\varepsilon (1-z)^2 + v_q z^{2 - \beta} (1-z) +
v_{qq}z^{2.5 - \beta}](1-z)^{\Delta \alpha}\; ,
\; G_{uu}^{\Lambda} =(1-z)G_{ud}^{\Lambda} \;
\end{equation}
\begin{equation}
G_{d,SJ}^{\Xi^-} = a_N z^{\beta} [v_0\varepsilon (1-z)^2 +
v_q z^{2 - \beta} (1-z)] (1-z)^{2\Delta \alpha}\; ,
\; ~G_{u,SJ}^{\Xi^-}=(1-z)G_{d,SJ}^{\Xi^-}\; ,
\end{equation}
\begin{equation}
G_{SJ}^{\Omega} = a_N v_0\varepsilon z^{\beta}
(1-z)^{2+3\Delta \alpha} \; .
\end{equation}

The factor $z^{\beta}$ is really $z^{1-\alpha_{SJ}}$. As for the factor
$z^{\beta} z^{2 - \beta}$ of the second term it is $2(\alpha_R - \alpha_B)$
\cite{KTM}. For the third term we have added an extra factor $z^{1/2}$.
The values $v_0$, $v_q$ and $v_{qq}$ were taken from \cite{ACKS}.

\par

The secondary baryon consists of the SJ together with two valence and one
sea quarks (Fig.~2a), one valence and two sea quarks (Fig.~2b) or three
sea quarks (Fig.~2c). The fraction of the incident baryon energy carried
by the secondary baryon decreases from a) to c), whereas the mean
rapidity gap between the incident and secondary baryon increases.
The diagram 2b has been used for the description of baryon number transfer
in QGSM \cite {KTM}. It describes also the fast pion production by a
diquark.

The probability to find a comparatively slow SJ in the case of Fig. 2c
can be estimated from the data on $\bar{p}p$ annihilation into
mesons (see Figs. 1c, d). This probability is known experimentally only at
comparatively small energies where it is proportional to
$s^{\alpha_{SJ}-1}$ with $\alpha_{SJ} \sim 0.5$.
However, it has been argued \cite{14r} that the annihilation cross section
contains a small piece which is independent of $s$ and thus $\alpha_{SJ}
\sim1$.

The main purpose of this paper is the determination of the contribution of
the graph in Fig.~2c to the diquark fragmentation function. Its magnitude
is proportional to a coefficient which will be denoted by $\varepsilon$.

Note that string-junction (as well as strings) has a nonperturbative
origin in QCD and at present it is impossible to determine $\alpha_{SJ}$
from QCD theoretically. Thus we treat $\alpha_{SJ}$, $\varepsilon$ and
$a_N$ as phenomenological parameters, which should be determined from
experimental data with the additional condition of baryon number
conservation.


\vskip 0.9 truecm
\noindent{\bf 3. COMPARISON WITH THE DATA}
\vskip 0.5 truecm

The mechanism of the baryon charge transfer via SJ without valence
quarks (Fig.~2c) was accounted for in previous paper \cite{ACKS},
where the value $\alpha_{SJ} = 0.5$ was used. Practically all
existing data at comparatively low energies
($\sqrt{s} \sim 15 \div 30$~GeV), were described with the value
$\varepsilon = 0.05$. However, the ISR \cite{ISR} data for the yields
of protons and antiprotons separately, as well as their differences are
described quite reasonably by QGSM with $\varepsilon =0.2$. The same
value of $\varepsilon =0.2$ allows one to describe HERA \cite{H1} data
on $\bar{p}/p$ asymmetry. This confirm the result \cite{Bopp} that the
$\bar{p}/p$ asymmetry measured at HERA can be obtained by simple
extrapolation of ISR data. It is necessary to note, that the systematic
errors in \cite{ISR} are of the order of 30~\%, so the value
$\varepsilon = 0.05$ can not be excluded. HERA data are preliminary and
have rather large errors. However, now the RHIC data on the $\bar{p}/p$
ratios in $pp$ collisions at $\sqrt{s}$ = 200 GeV appear \cite{RHIC} and
these data are in agreement with HERA data.

Some part of disagreement in the values of $\varepsilon$ parameter at
different energies can be connected with phase space effects. In the
process of Fig. 2c , as a minimum, two additional mesons $M$ should be
produced in one of the strings, that can give an additional suppression
\cite{TMS} of the process Fig. 2c in comparison with, say Fig. 2a.
However the 4 times difference in the values of $\varepsilon$ parameter
obtained in \cite{ACKS} in different energy regions seems to be too large
for phase space suppression.

Another possibility to explain this difference is that the value of the
intersept $\alpha_{SJ} = 0.5$ in \cite{ACKS} was taken too small. Really,
the SJ contribution to the inclusive cross section of secondary baryon
production at the rapidity distance $\Delta y$ from the incident particle
can be estimated as
\begin{equation}
d \sigma_B /d y \sim a_B \varepsilon e^{(\alpha_{SJ} -1) \Delta y} \; ,
\end{equation}
$a_B = a_N v_i$. So the increase of the effective value of $\varepsilon$
with the energy, i.e. with $\Delta y$ can be consider as a signal to
increase the value of $\alpha_{SJ}$.

In the presented paper we found the solution of the problem with the
parameters

\begin{equation}
\alpha_{SJ} = 0.9 \; ,\; \varepsilon = 0.024 \; ,\; a_N = 1.33 \;
{\rm (low \; energies)} \; ,\; a_N = 1.29\; {\rm (high \; energies)} \; ,
\end{equation}
and the values $v_i$ in Eqs. (6)-(9) were taken from quark combinatorics
\cite{ACKS,CS}

The quality of the description of the experimental data with these
parameters is practically the same as in \cite{ACKS}. As an example
we present in Fig.~3 the inclusive spectra of secondary protons in
$pp$ collisions at lab. energies 100 and 175 GeV \cite{Bren}.

In Fig.~4 we show the data \cite{ait1} on the asymmetry of strange baryons
produced in $\pi^-$ interactions\footnote{These data were obtained
from pion interactions on a nuclear target where different materials were
used in a very complicated geometry. We assume that the nuclear effects
are small in the asymmetry ratio (12), and compare the pion-nucleus
data with calculations for $\pi^-p$ collisions.} at 500 GeV/c. The
asymmetry is determined as
\begin{equation}
A(B/\bar{B}) = \frac{N_B - N_{\bar{B}}}{N_B + N_{\bar{B}}}
\end{equation}
for each $x_F$ bin.

The theoretical curves for the data on all asymmetries calculated with
parameters (11) are in reasonable agreement with the data. Sometimes this
agreement is even better than in \cite{ACKS}.

In the case of $\Omega/\bar{\Omega}$ production we predict a non-zero
asymmetry in agreement with experimental data. Let us note that the last
asymmetry is absent, say, in the naive quark model because $\Omega$ and
$\bar{\Omega}$ have no common valence quarks with the incident particles.

Preliminary data on $p/\bar{p}$ asymmetry in $ep$ collisions at HERA were
presented by the H1 Collaboration \cite{H1}. Here the asymmetry is defined
as
\begin{equation}
A_B = 2 \frac{N_p - N_{\bar{p}}}{N_p + N_{\bar{p}}} \; ,
\end{equation}
i.e. with an additional factor 2 in comparison with Eq. (12). The
experimental value of $A_B$ is equal to $8.0 \pm 1.0 \pm 2.5$ \%
\cite{H1} for secondary baryons produced at $x_F \sim 0.04$ in the
$\gamma p$ c.m. frame. In QGSM the hadron structure of photon is
considered as ($\pi^+ + \pi^-$)/2 \cite{LSS}. Such approach with
parameters (11) leads to the value $A_B$ = 9.9~\%, in agreement with
the data. The experimental value $A_B$ was predicted in \cite{Kop}
using $\alpha_{SJ} = 1$ that is rather close to our choice (11).

The baryon charge transferred to large rapidity distances can be
determined by integration of Eq. (10), so it is of the order of
\begin{equation}
<n_B> \sim a_B \varepsilon /(1 - \alpha_{SJ}).
\end{equation}
It is clear that the value $\alpha_{SJ} = 1$ should be excluded due to the
violation of baryon number conservation at very high energies. At the same
time the values of $\alpha_{SJ}$ very close to unity are possible. If the
essential part of the initial baryon charge is transferred to large
rapidity distances, the altitude of secondary baryon spectra in the proton
fragmentation region (large $x_F$) should be decreased. This leads to
violation \cite{Sh2} of Feynman scaling at very high energies. For
example, we predict decrease of the secondary neutron multiplicity
with $x_F > 0.28$ from 0.324 to 0.27 for energy region
$\sqrt{s} = 20 \div 200$ GeV. The experimental estimation of this effect
\cite{ZEUS} is significantly larger, about two times.

The RHIC $pp$ data \cite{RHIC} on the ratio of $\bar{p}/p$ at low values
of c.m. rapidity also are described reasonably with parameters (11) as one
can see in Fig. 5.

Some our predictions at $\sqrt{s} = 200$ GeV for the
antihyperon/hyperon production asymmetries in $\gamma p$ and
$\bar{B}/B$ ratios in $pp$ collisions are presented in Table 1.

\begin{center}
{\bf Table 1}
\end{center}
\vspace{15pt}
The predicted values of the antihyperon/hyperon production asymmetries
Eq. (13) in $\gamma p$ collisions and $\bar{B}/B$ ratios in $pp$
collisions, both at $\sqrt{s} = 200$ GeV.
\begin{center}
\vskip 12pt
\begin{tabular}{|c|c|c|}\hline

Hyperons  &  $A_B$       &  $\bar{B}/B $  \\   \hline

 $\bar{\Lambda},\Lambda$ &  10.8 \%   & 0.77   \\

 $\bar{\Xi},\Xi$         &  6.5 \%   & 0.82   \\

$\bar{\Omega},\Omega$    & 12.0 \%   & 0.81    \\

\hline
\end{tabular}
\end{center}

\vskip 1.0cm
\vskip 0.9 truecm
\noindent{\bf 5. CONCLUSIONS}
\vskip 0.5 truecm

We presented the role of string junction diffusion for the baryon charge
transfer over large rapidity distances. Without this contribution shown in
Fig. 2c the data for hyperon/antihyperon asymmetries (Fig. 4),
proton/antiproton asymmetry \cite{H1} and $\bar{p}/p$ ratios at RHIC
(dashed curve in Fig.~5) are in total disagreement with the data.

It is necessary to note that value of $\varepsilon$ parameter in (11) was
taken from the normalization to one experimental point, parameter
$\alpha_{SJ}$ was found from energy (or rapidity) dependence of the
observed effects and the value $a_N$ was calculated from the condition of
baryon number conservation. The results of calculations for another
processes which are also sensitive to string junction diffusion of
baryon charge are practically the same as in \cite{ACKS}, so we do not
present them. It is necessary to note that the existing experimental data
are not enough for determination of the SJ parameters with the needed
accuracy.

This paper was supported by DFG grant GZ: 436 RUS 113/771/1-2 and, in
part, by grants RSGSS-1124.2003.2 and PDD (CP) PST.CLG980287.

\newpage

{\bf Figure Captions}

\vskip 1cm

Fig. 1. Cylindrical diagram corresponding to the one-pomeron exchange
contribution to elastic $\bar{p}p$ scattering (a) and its cut which
determines the contribution to inelastic $\bar{p}p$ cross section (b)
(string junction is indicated by a dashed line). The diagram for elastic
$\bar{p}p$-scattering with SJ exchange in the $t$-channel (c) and
its $s$-channel discontinuity (d) which determines the contribution to
annihilation $\bar{p}p$ cross section.

Fig. 2. Three different possibilities of secondary baryon production in
$pp$ interactions via diquark $d$ fragmentation: string junction together
with two valence and one sea quarks (a), together with one valence and
two sea quarks (b), together with three sea quarks (c).

Fig. 3. The spectra of secondary protons (a)
in $pp$ collisions at 100, 175 GeV/c \cite{Bren} and their description
by QGSM.

Fig. 4. The asymmetries of secondary  $\Lambda/\bar{\Lambda}$ (a),
$\Xi^-/\Xi^+$ (b) and $\Omega/\bar{\Omega}$ (c) in $\pi^-p$ collisions
at 500 GeV/c \cite{ait1} and its description by QGSM. 

Fig. 5. The ratios of secondary antiproton/proton production in
$pp$ interactions at $\sqrt{s} = 200$ GeV (points and solid curve).
The calculated result with $\varepsilon = 0$ is shown by dashed curve.

\end{document}